\documentclass[preprint,showpacs,preprintnumbers,amsmath,amssymb]{revtex4-2}
\usepackage[dvipdfmx]{graphicx}

\usepackage{graphicx}
\usepackage{dcolumn}
\usepackage{bm}
\usepackage{color}
\usepackage{rotating}
\usepackage{ulem} 

\begin{document}

\title{Universal scaling of the specific heat  in  $S=1/2$ quantum kagome antiferromagnet herbertsmithite}

\author{H.\,Murayama$^1$}
\author{T. Tominaga$^1$}
\author{T. Asaba$^1$}
\author{A. de Oliveira Silva$^1$}
\author{Y. Sato$^{1}$}
\author{H. Suzuki$^1$}
\author{Y. Ukai$^1$}
\author{S. Suetsugu$^1$}
\author{Y. Kasahara$^1$} 
\author{R. Okuma$^2$}
\author{I. Kimchi$^3$}
\author{Y.\,Matsuda$^1$}

\affiliation{$^1$ Department of Physics, Kyoto University, Kyoto 606-8502 Japan}
\affiliation{$^2$ Clarendon Laboratory, University of Oxford, Parks Road, Oxford OX1 3PU, UK} 
\affiliation{$^3$ School of Physics, Georgia Institute of Technology, Atlanta, GA 30332, USA}



\begin{abstract}
{ 
Despite tremendous investigations, a quantum spin liquid state realized in spin-1/2 kagome Heisenberg antiferromagnet remains largely elusive.  In herbertsmithite ZnCu$_3$(OH)$_6$Cl$_2$,  a quantum spin liquid candidate on the perfect kagome lattice, precisely characterizing the intrinsic physics of the kagome layers is extremely challenging due to the presence of  interlayer
Cu/Zn antisite disorder within its crystal structure. 	 Here we measured the specific heat and thermal conductivity of single crystal herbertsmithite in magnetic fields with high resolution. 
Our results are highlighted by the excellent scaling collapse of the intrinsic magnetic specific heat contribution arising from the kagome layers as a function of $T/H$ (temperature/magnetic field).  
In addition, no residual linear term in the thermal conductivity $\kappa/T(T\rightarrow 0)$  is observed in zero and applied magnetic fields,  indicating the absence of itinerant gapless excitations.  These results suggest a new picture for a quantum spin liquid state of the kagome layers of herbertsmithite, wherein  localized orphan spins arise and interact with random exchanges in conjunction with a non-itinerant quantum spin liquid. }
\end{abstract}
\maketitle

	A quantum spin liquid (QSL) is an exotic state of matter where quantum fluctuations obstruct the formation of long-range magnetic order even in the zero-temperature limit.  In QSLs, spins are quantum mechanically entangled over long distances without showing simple symmetry-breaking, and they can form fractionalized collective excitations.    For spin systems in dimensions higher than one, it is generally believed that frustrating interactions are required to stabilize the QSL states.    
Among this class of materials, the spin-1/2 two-dimensional (2D) kagome Heisenberg antiferromagnet with strong geometrical frustration has attracted considerable interest, as such a system is supposed to exhibit a QSL ground state\cite{Sachdev_QSLonKagome_1992}.  
However, understanding the nature of the kagome lattice has proved to be one of the most vexed issues in the quantum spin systems.  In fact, despite tremendous research efforts, the ground state of the QSL in the kagome system remains unknown
\cite{Ran_GutzwillerProjection_U1Dirac_2007, Singh_SeriesExpansion_VBC_2007, Hermele_GutzwillerProjection_AlgebraicSL_2008, Jiang_DMRG_gapped_2008, Ran_DiracSL_LandauLevel_2009,Gotze_CoupledCluster_QSL_2011, Yan_DMRG_gappedSL_2011, Jiang_DMRG_topologicalSL_2012, Depenbrock_DMRG_Z2_2012, Nishimoto_Z3SL_Gapped_2013, Iqbal_DMRG_DiracSL_2013, clark_StripedSL_2013, He_DMRG_U1Dirac_2017, Lauchli_Lancoz_Z2SL_2019}. 
Currently, the most promising candidates are a gapped spin liquid with a $Z_2$ topological order
\cite{Jiang_DMRG_gapped_2008, Jiang_DMRG_topologicalSL_2012, Depenbrock_DMRG_Z2_2012, Lauchli_Lancoz_Z2SL_2019}
 and a gapless $U(1)$ Dirac  spin liquid
\cite{Ran_GutzwillerProjection_U1Dirac_2007, Ran_DiracSL_LandauLevel_2009, Iqbal_DMRG_DiracSL_2013, He_DMRG_U1Dirac_2017}.  

Among kagome quantum magnets, herbertsmithite ZnCu$_3$(OH)$_6$Cl$_2$  has been most extensively studied as a canonical candidate for bearing a QSL  state\cite{Mendels_review_2010, Norman_review_2016}.
The crystal consists of  2D perfect kagome planes nearly fully occupied with Cu$^{2+}$ ions,  and a kagome lattice of spin-1/2 nearest-neighbor Heisenberg antiferromagnetic interactions is realized.   The crystal structure of herbertsmithite contains interlayer sites  primarily occupied by nonmagnetic Zn$^{2+}$ ($S=0$) (Figs.\,1(a) and 1(b))\cite{Shores_Cyrstalgrowth_2005, Han_Crystalgrowth_2011}.   Some interlayer Zn sites are replaced by  Cu, which induces antisite disorder within its crystal structure\cite{Freedman_Xray_2010}.  In contrast to the other kagome candidates,   herbertsmithite does not exhibit magnetic ordering  down to the lowest measured temperatures, despite large exchange interaction ($J\approx$180\,K) in the kagome layers
\cite{Ofer_NMR_SR_2006, Mendels_SR_Cudoped_2007, Bert_chi_lowtemp_2007}.
The inelastic neutron scattering (INS) measurements revealed that excitations are dominated by an unusual broad continuum, which has been considered to be a signature of the fractional spinon excitations in the QSL. \cite{Han_Neutron_2012}.

A key question is to understand the intrinsic physics of the kagome layers in herbertsmithite, particularly magnetic and thermodynamic properties.  Despite intensive experimental investigations, however,  precisely characterizing the low-energy excitations within the kagome layers is extremely challenging.  In fact, recent studies have invoked different aspect of this compound, i.e. the interlayer Cu/Zn antisite disorder has a significant impact on these properties
\cite{Olariu_NMR_2008, Imai_NMR_2008, Nilsen_Neutron_LowEnergy_2013, Han_Neutron_2016, Fu_NMR_2015, Khuntia_NMR_2020}.   
Although nuclear magnetic resonance (NMR)\cite{Fu_NMR_2015,Khuntia_NMR_2020}
and INS experiments\cite{Han_Neutron_2016} have been performed by several groups, whether the spin excitations are gapped or gapless is still controversial.   Furthermore,  interpretation of the most fundamental thermodynamic quantities, such as specific heat\cite{Helton_C_powder_2007, Devries_C_powder_2008, Han_C_single_2012, Shaginyan_C_single_scaling_2012, Han_C_18T_single_2014}
 and magnetic susceptibility\cite{Bert_chi_lowtemp_2007, Rigol_chi_TheorywithDM_2007, Helton_chi_scaling_2010, Bernu_chi_HighTSeriesExpasion_2015, Hotta_chi_gapless_2018}, 
remains largely elusive.    It has been suggested that intrinsic specific heat of kagome layers may be seriously masked by the contribution from the antisite disorder, which dominates the total specific heat.   The magnetic susceptibility exhibits a diverging Curie-like tail, suggesting that some of the Cu spins act as weakly coupled impurities\cite{Bert_chi_lowtemp_2007}.

Recently, a new mechanism for unusual features in thermodynamic quantities of quantum spin systems has been  introduced\cite{Kimchi_DMRG_VBS_2018, kimchi_scaling_2018}.     
For this scenario,  orphan spins are induced by randomness or disorder, and form random singlets.  In some quantum spin systems,  including valence bond solids and  QSLs with sufficient disorder,  it has been proposed that low temperature specific heat $C(H,T)$ in temperature $T$ and magnetic field $H$ exhibits $T/H$ collapse, showing universal scaling features.     The universal scaling appears as a result of a broad distribution of antiferromagnetic exchange interactions, which is a driving force of the formation of such random singlets.  It has been pointed out that this scaling relation may hold in herbertsmithite\cite{kimchi_scaling_2018}.   In this scenario, observed specific heat could arise from the kagome layers, in contrast to previous interpretations.  However, it is premature to judge the validity of this scaling collapse because of the following reasons.  First of all,  extra contributions to specific heat, such as phonon term and contribution arising from interlayer Cu/Zn antisite disorder,  are not excluded from the measured $C(T)$ and hence it must be checked whether the scaling law is valid or not after subtracting these contributions. Indeed, distinct deviations from the scaling law can be seen at some $T/H$ range.   
In addition, the scaling assumes that the magnetic excitations are localized, but it is open whether the specific heat contains the itinerant magnetic excitations.   Moreover, the measurements were performed on powdered sample\cite{Helton_C_powder_2007}, where the magnetic anisotropy of the specific heat is smeared out.  

\begin{figure}[t]
	\centering
	\includegraphics[width=1\linewidth]{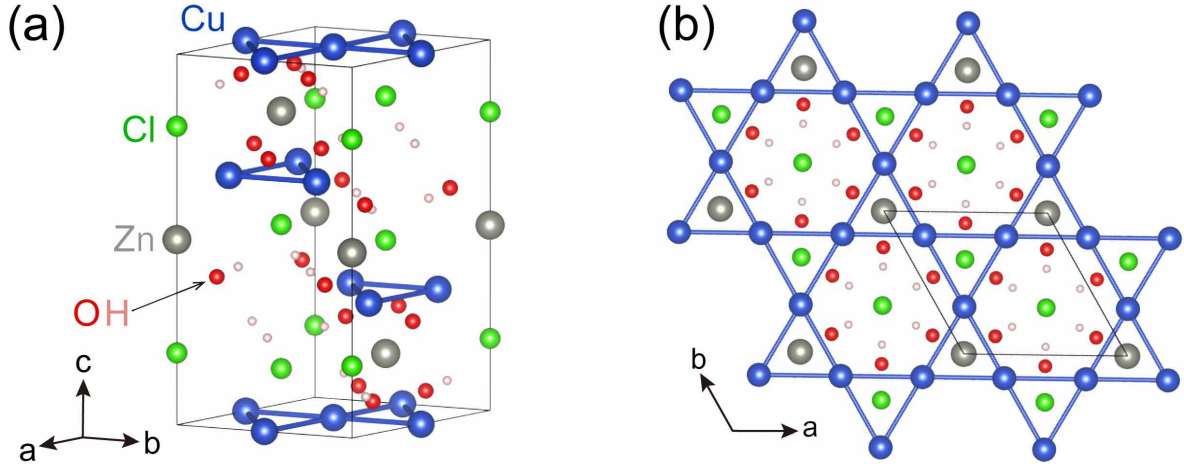}
	\caption{\label{figure1}(a) The unit cell of Herbertsmithite  ZnCu$_3$(OH)$_6$Cl$_2$. ABC-stacked kagome layers composed of Cu are separated by Cl, Zn and OH. Zn forms an octahedron with oxygens. (b) The top view of a kagome layer. Spins on Cu sites are coupled through Cu-O-Cu super-exchange interaction. }
\end{figure}

Thus, examining the validity of the scaling relation of the specific heat is crucially important to understand the intrinsic thermodynamic properties of the kagome layers.  While specific heat contains both localized and itinerant excitations, thermal conductivity only detect the itinerant contribution.  Therefore the combined results of specific heat and thermal conductivity provide pivotal information on the low energy excitations.    In this Letter, we measured the specific heat and thermal conductivity $\kappa$ with high accuracy on single crystals of herbertsmithite.    
The most important finding is that the intrinsic magnetic contribution of the specific heat in the kagome layers exhibits excellent scaling collapse for $T/H$.  This implies that the specific heat in the kagome layers is governed by localized orphan spins that form random singlets. 

High-quality single crystals are prepared by recrystallization in a three-zone furnace after the prereaction\cite{Han_Crystalgrowth_2011}.
For the accurate measurements of the specific heat, we used the long relaxation method\cite{Carrington_LongRelaxation_2007} on a single crystal ($1.5\times1.2\times0.5$\,mm$^3$, 5.9\,mg).
The in-plane thermal conductivity was measured on a single crystal with dimensions $1.5\times0.7\times0.2$\,mm$^3$ cut from the crystal used for the specific heat measurements by the standard steady-state technique. For both measurements, the magnetic field was applied perpendicular to the 2D plane ({\boldmath $H$}$\parallel c$).

\begin{figure}[t]
	\centering
	\includegraphics[width=0.65\linewidth]{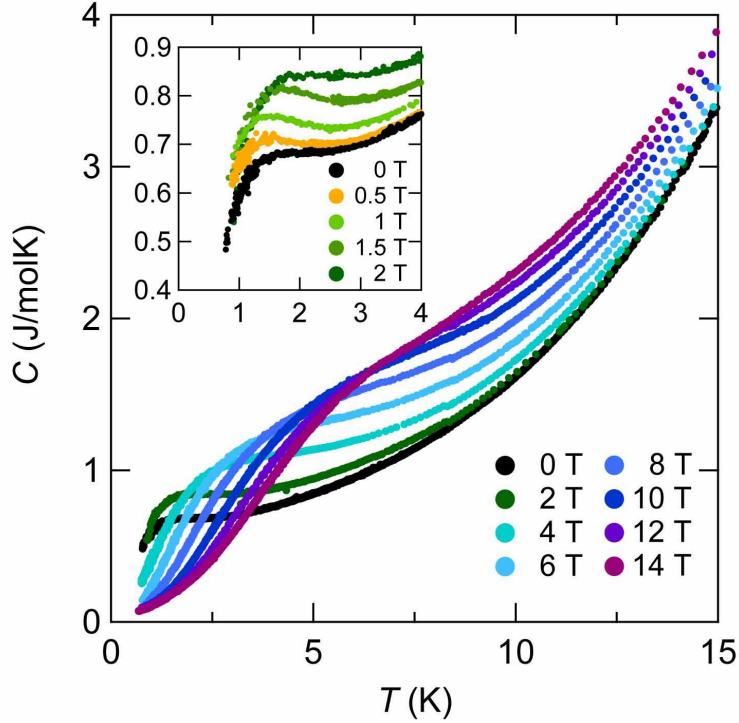}
	\caption{\label{figure2}Temperature dependence of the specific heat in zero and applied magnetic field up to 14\,T ({\boldmath $H$}$\parallel c$).  The inset shows the specific heat in the low temperature regime at low fields. }
\end{figure}

  The inset of Fig.\,2 depicts the $T$-dependence of $C$ in zero and at low fields in the low temperature regime.    In zero field, as the temperature is increased,  $C$ first increases steeply,  showing a shoulder structure, and then increases upwardly.     In magnetic fields, the shoulder structure is pronounced,  resulting in the broad maximum around 1.5\,K.  The main panel of Fig.\,2 depicts the temperature dependence of  $C$ up to 15\,K in zero and finite magnetic fields.  At higher fields above 4\,T,  $C$ increases monotonically with elevating temperature, showing a hump structure.  The hump temperature increases with increasing field.    For all fields, $C$ increases upwardly in the high temperature regime above $\sim$10\,K.
As shown in Fig.\,S1(a),  $C/T$ increases in proportion to $T^2$ at high temperatures for all fields and the field-dependent data overlap with each other after vertically shifting.   
  This cubic temperature dependence of $C$ is attributed to the acoustic phonon contribution; $C_{ph} = \beta_{ph} T^3$, where $\beta_{ph}$ is the Debye coefficient.  From the fitting,  we obtain $\beta_{ph}=6$\,-\,$7\times 10^{-4}$\,J/mol\,K$^4$, which corresponds to the Debye temperature of $375$\,-\,$395$\,K.   
   
As depicted in Fig.\,3, $\kappa/T$ in zero field increases almost linearly with temperature, but if we extrapolate $\kappa/T$ to zero temperature simply assuming $T$-linear dependence, $\kappa/T$ has a negative intercept.  In the inset of Fig.\,3, $\kappa/T$ is plotted as a function of $T^2$.   Obviously, $\kappa/T$ increases with decreasing slope.   The results of Fig.\,3 and its inset, which plots $\kappa/T$ vs. $T^2$, indicate that $\kappa/T$ depends on $T$ as $\kappa/T\propto T^\alpha$ with $1\alt\alpha<2$.   The best fit is obtained by $\alpha=1.3$, as shown by the dotted line in the inset of Fig.\,3.   These results indicate that the residual linear term of the thermal conductivity, $\kappa_0/T \equiv \kappa/T(T\rightarrow 0)$, is vanishingly small, if present at all.  This provides evidence for the absence of gapless itinerant excitations\cite{Yamashita_dmit_2010, Yamashita_dmit_2020, Murayama_TaS2_2020}.   The red open circles in Fig.\,3 and its inset show $\kappa/T$ in magnetic field of  $\mu_0H$=14\,T.  In stark contrast to large field-dependent specific heat shown in Fig.\,2,  the magnetic field has no influence on the thermal conductivity.  This field independent $\kappa$ indicates that thermal conduction is dominated by phonon contribution, $\kappa\approx \kappa_{ph}$ (see Supplementary for details).  Moreover, as the magnetic field suppresses the spin-phonon scattering by polarizing spins, the phonon mean free path increases with magnetic field.  The present results, therefore,  indicate negligibly small spin-phonon coupling.

 \begin{figure}[t]
	\centering
	\includegraphics[width=0.65\linewidth]{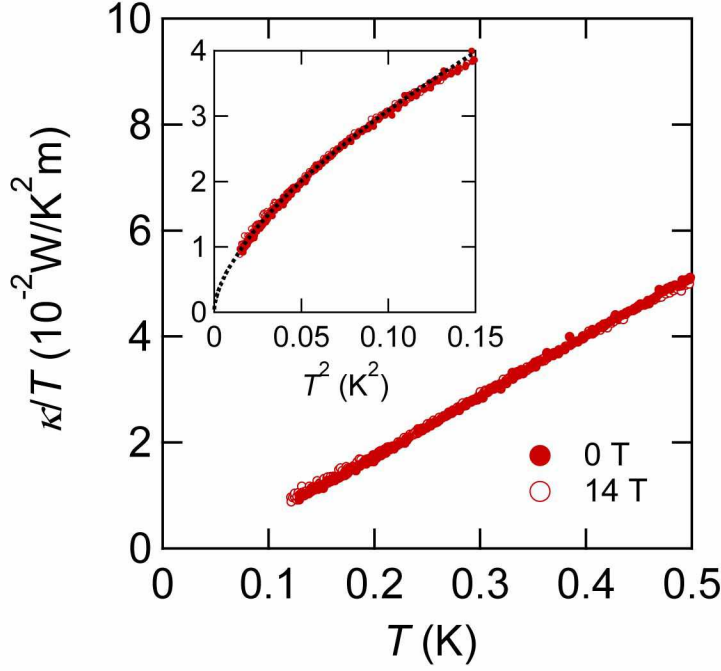}
	\caption{\label{figure4}Thermal conductivity divided by temperature, $\kappa/T$  plotted as a function of $T$. Red filled and open circles represent the data in zero field and applied field of $\mu_0H=$14\,T  ({\boldmath $H$}$\parallel c$), respectively.  The inset shows $\kappa/T$ vs. $T^2$.    A dashed line represents $\kappa/T\propto T^{1.3}$.}
\end{figure}

Having established the absence of itinerant gapless quasiparticle excitations, we analyze the temperature dependence of the specific heat in more detail.    We point out that the observed shoulder structure in zero field and  broad maximum at weak field of $C$ can be attributed to the two level Schottky specific heat as
\begin{equation}
	C_{Sch}=A_{Sch}\left[\frac{\Delta(H)}{k_BT}\right]^2\exp\left[-\frac{\Delta(H)}{k_BT}\right], 
	\label{Eq:Sch}
\end{equation}
where $A_{Sch}$ is a constant that is determined by the number of two-level systems. $\Delta(0)=g\mu_BH_0$  is the energy of the excited level, where $H_0$ is the magnetic field characterized by the crystal electric field and  $g$ is the electron $g$-factor assumed to be 2.  
We assume that the specific heat contains magnetic contribution $C_{mag}$ and is given by $C=C_{ph}+C_{Sch}+C_{mag}$.     Figure\,S2(a) depicts $C$ in zero field below 4\,K  where phonon contribution is negligibly small.    We try to fit $C(T)$ at low temperatures by assuming power-law temperature dependence of $C_{mag}=A_{mag}T^{1-\eta}$, where $A_{mag}$ and $\eta$ are constants.  As shown in Fig.\,S2(a), $C(T)$ is well fitted by $A_{Sch}=0.472(6)$, $H_0=1.814(8)$, $A_{mag}=0.350(3)$ and $\eta=0.581(4)$. 
Surprisingly, as shown in Fig.\,4(a), $C(T)$ in zero field is excellently fitted by these three contributions up to 15\,K.   In the fitting,  we used $\beta_{ph}=6.78\times10^{-4}$\,J/mol\,K$^4$.  We note that the two-level Schottky specific heat likely arises from the interlayer Cu/Zn antisite disorder.  The number of two-level obtained from $A_{Sch}$ indicates that nearly 5\% of Zn site is replaced by Cu.   This value is  nearly 1/3 of that reported by NMR\cite{Fu_NMR_2015} and resonant x-ray diffraction\cite{Freedman_Xray_2010} measurements.  The reason for this discrepancy between the measurements is not clear.

\begin{figure}[t]
	\centering
	\includegraphics[width=0.6\linewidth]{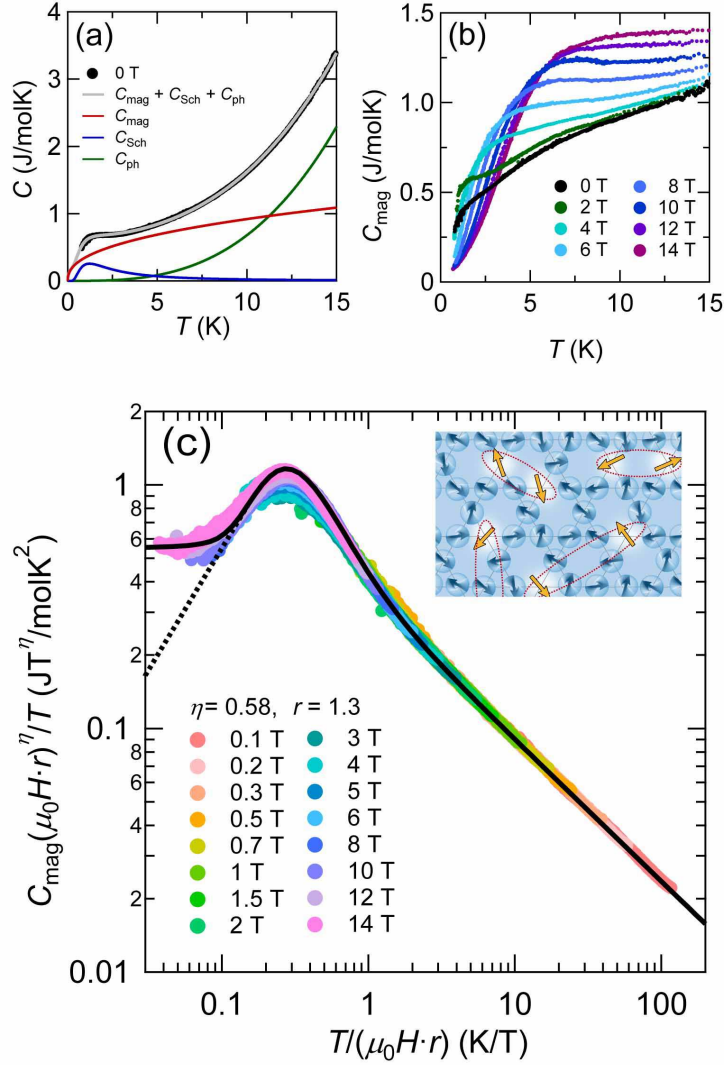}
	\caption{\label{figure3}(a) Specific heat  $C$ in zero field plotted as a function of $T$.   The gray line represents the sum of phonon (green line) , Schottky (blue) and magnetic (red) contributions.     (b) Temperature dependence of magnetic contribution $C_{mag}$ in zero and applied magnetic fields obtained by subtracting $C_{Sch}$ and $S_{ph}$ from total $C$.    (c)  Scaling relationship of magnetic  contribution of the specific heat; $C_{mag}(\mu_0Hr)^{\eta}/T$ plotted as a function of $T/(\mu_0Hr)$. Excellent scaling is observed with $r=1.3$.   Solid black line indicates the scaling function obtained by the fit to $F_0$ in Eq.\,(\ref{Eq:scaling}) (cf. dotted black line showing $F_1$).}
\end{figure}

To confirm the validity of the present analysis, we fit the low temperature data at low fields, where $C(T)$ exhibits a broad maximum shown in the inset of Fig.\,2.   In the fitting, we fixed $A_{Sch}$.   As shown in Figs.S2(b)-(j), $C(T)$  is again well fitted by Schottky contribution and power-law dependent $C_{mag}$ in this temperature range.  The field dependence of $A_{mag}$, $\eta$ and $\Delta$ are shown in Figs.\,S2(k)-(m).   In the two-level Schottky model,   the excited energy level depends on $H$ as  
\begin{equation}
\Delta(H)=g\mu_B\sqrt{H^2+H_0^2}.
\label{Eq:gap}
\end{equation}
In Fig.\,S2(m), the red line represents Eq.\,(\ref{Eq:gap}).  As shown in Fig.\,S2(m), $\Delta(H)$ determined by the fitting shown in Figs.\,S2(a)-(j) reasonably coincides with the results of Eq.\,(\ref{Eq:gap}).  The small deviation may be due to the assumption of the  power law dependent $C_{mag}$ in magnetic fields.   Indeed, in contrast to the excellent fitting in zero field $C(T)$ in the whole temperature regime,  $C(T)$ starts to deviate at high temperatures when magnetic fields are applied.     In what follows,  to extract $C_{mag}$,   we use $C_{Sch}$ calculated from  Eq.\,(\ref{Eq:Sch}) with $\Delta(H)$ given by Eq.\,(\ref{Eq:gap}).  Moreover,  $C_{ph}$ is calculated by assuming field-independent $\beta_{ph}=6.78\times10^{-4}$\,J/mol\,K$^4$, which is justified by the field-independent thermal conductivity.

Figure\,4(b) depicts the $T$-dependence of $C_{mag}$ obtained by subtracting $C_{ph}$ and $C_{Sch}$.   Recently,  new theoretical studies of the role of quenched disorder in quantum paramagnetic states including QSL have been proposed\cite{Kimchi_DMRG_VBS_2018, kimchi_scaling_2018}.    In this picture,  as shown schematically in the inset of  Fig.\,4(c), the majority of spin-1/2 sites form a quantum paramagnetic state such as a spin liquid, while a small fraction of sites host nucleated ``orphan spins," which need not be microscopic defects but rather could be emergent quantum objects that can arise from a competition of disorder and frustration.  An orphan spin can couple with another orphan spin to form a singlet state.  The exchange energies between these orphan spins vary randomly, with an exponential dependence on their distance, which leads to a formation of singlets with random energy gaps whose distribution is exponentially broad.  This broad distribution includes singlets with arbitrarily small energy gaps.   This model predicts that $C_{mag}$ arising from the localized spin excitations collapses into a single curve of form,
\begin{equation}
	\frac{C_{mag}(H,T)}{T}\sim \frac{1}{H^{\eta}}F_q(T/H),
	\label{Eq:scaling}
\end{equation} 
where $F_q(X)$ is a scaling function which is determined by  the energy distribution of the random singlets;  
\begin{eqnarray}
	F_q(X) \sim \left\{
	\begin{array}{ll}
		X^q  &X \ll 1  \\
		X^{-\eta}(1+c_0/X^2) &X \gg 1.
	\end{array}\right.
\end{eqnarray}
Here $q=1$ and $q=0$ correspond to the case with and without Dzyaloshinskii-Moriya interactions, respectively, $\eta$ is a non-universal exponent, $0\leq \eta \leq1$, that characterizes the probability distribution of antiferromagnetic exchange energies $P(J)\sim J^{-\eta}$.  The spins with exchange $J <k_BT$ behave as free spins and $C(T)$ shows power-law dependence on $T$ as $C\propto T^{1-\eta}$.   When $\eta\neq 0$,  $F_q(X)$  increases with $X$, peaks   and  decreases at large $X$, saturating for $q=0$. It should be stressed that the power-law temperature dependence of $C(T)$ in zero field is consistent with this scaling and $\eta$ is uniquely determined.

In Fig.\,4(c),  $C_{mag}(\mu_0Hr)^{\eta}/T$ is plotted as a function of $T/(\mu_0Hr)$.  In this plot, the fitting parameter is  $r=g_o/2$, where $g_o$ is the effective $g$-factor of orphan spin.   We find that $C_{mag}$ at all fields collapse into  a single curve with $r$=1.3.    The scaling function obtained by the fit to $F_0$ in Eq.\,(\ref{Eq:scaling})  is shown by the solid black line.  Thus $C_{mag}$ exhibits an excellent  scaling collapse with a universal scaling function. 
At $T/(\mu_0Hr)<0.1$, $C_{mag}(\mu_0Hr)^{\eta}/T$ becomes constant, suggesting $q=0$. To confirm this, it is necessary to measure the specific heat precisely at $T/(\mu_0Hr)\ll0.1$, where the specific heat is dominated by the nuclear Schottky anomaly which makes it difficult to evaluate the magnetic excitations\cite{kimchi_scaling_2018}. 

The scaling law provides several pieces of important information on the QSL state in herbertsmithite. 
The result suggests that the specific heat contains substantial contributions from the frustrated kagome layers. In addition, the quantum fluctuations in kagome layers sensitively respond to randomness.    In herbertsmithite, randomness is likely attributable to the exchange bond disorder in the perfect kagome lattice, which is caused by the interlayer Cu/Zn antisite disorder.  In fact,  the Jahn-Teller effect causes a local structural distortion around the Cu defects in the Zn planes\cite{Lee_Cudoped_Neutron_2007},  which gives rise to the exchange bond disorder of Cu ions in the kagome layer just above and below the defects.   
If antisite disorder with Cu spins on interlayer sites are reasonably decoupled from the kagome layers, their density can be estimated from their Schottky anomaly contribution to specific heat, giving 5\% of Zn sites (equivalently a density of 1.7\% of overall Cu sites). However, the entropy associated with the scaling part of $C(T)$, estimated by $S_{mag}= (1/R \ln 2) \times \int^{T}_{0}\frac{C'_{mag}(T')}{T'}dT'$, reaches much larger value of 0.05 (per Cu site) already by $T =1$\,K, rising to 0.1 by $T = 5$\,K (Fig. S3). It therefore seems inconsistent to attribute the specific heat scaling purely to Zn/Cu defect spins (or, at the very least, such spins cannot be considered as independent variables). 

We point out that the orphan spins in the kagome layers also largely contribute to the magnetic susceptibility, which exhibits a diverging behavior with decreasing temperature, although the quantitative analysis is difficult compared with the specific heat.     
Moreover, the QSL state of the herbertsmithite does not appear as a result of the destruction of the magnetically ordered ground state due to the randomness\cite{Kawamura_Randomness_ExactDiagonalization_2014, Kawamura_randomness_review_2019}.  Rather, the ground state of this system is the QSL state originated from the quantum fluctuations and frustration.  Randomness induces a fraction of orphan spins forming localized random singlets in the kagome layers.  Finally we note that a similar scaling relation of the specific heat has been reported in 1T-TaS$_2$, a triangular lattice system which may form spinon Fermi surface\cite{Murayama_TaS2_2020}.   Given that the present results of herbertsmithite provide strong support for similar $q=0$ scaling collapse in Ref.\cite{kimchi_scaling_2018}, taken together these results hint at a potentially universal feature for QSLs with weak randomness.  

In summary,  we measured the specific heat and thermal conductivity on single crystals of herbertsmithite.  Thermal conductivity reveals the absence of gapless itinerant excitations.    Our result is highlighted by an excellent scaling collapse for $T/H$ of the intrinsic magnetic contribution of the specific heat in the kagome layers.  These results demonstrate that the specific heat in the kagome layers is governed by localized orphan spins that form random singlets, which are surrounded by the quantum spin liquid. The present study provides vital information on how the quantum fluctuations respond to randomness due to quenched disorder, which is key for fundamental understanding the mysterious QSL states.

After completion of this work, we became aware of the Ref.\cite{Huang_C_k_2021}, which also discusses the scaling of the same compound. 

We thank H. Kawamura, Z. Hiroi, K. Totsuka for helpful discussions. This work was supported by Grants-in-Aid for Scientific Research (KAKENHI) (No. 18J10553, and No. 18H05227) and  JST CREST (No. JP-MJCR18T2).

\end{document}


\title{Supplementary}

%
%
%
%
%
%
%



\maketitle

\def\thesection{S\arabic{section}}
\def\theequation{S\arabic{equation}}
\def\thefigure{S\arabic{figure}}


\section{High temperature specific heat}
\begin{figure}[h]
	\centering
	\includegraphics[width=0.6\linewidth]{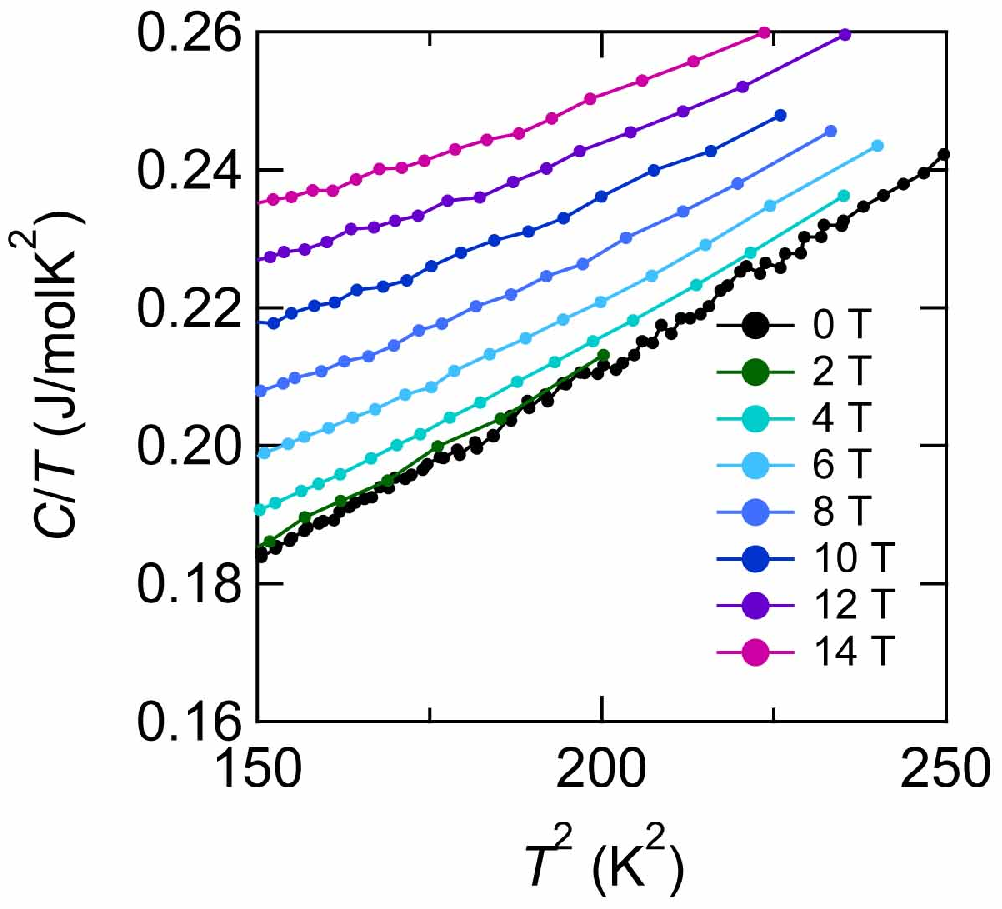}
	\caption{Specific heat divided by temperature $C/T$ vs. $T^2$ in magnetic field ({\boldmath $H$}$\parallel c$) in the high temperature regime. To determine the phonon contribution reliably, we measured the specific heat up to above 15\,K.}
\end{figure}

\section{Calculation of phonon mean free path}
At low temperatures,  $\kappa_{ph}$ is given by $\kappa_{ph}=\frac{1}{3}\beta_{ph} \langle v_s\rangle \ell_{ph}T^3$, where $ \langle v_s\rangle$ is the acoustic phonon velocity, and $\ell_{ph}$ is the effective mean free path of acoustic phonons.   When  $\ell_{ph}$ becomes comparable to the crystal size at very low temperatures (boundary limit),  $\ell_{ph}$ is approximately limited by the effective diameter of the crystal $d_{eff}=2\sqrt{wt/\pi}$, where $w$ and $t$ are the width and thickness of the crystal, respectively.     Using $\beta_{ph}=6.78\times10^{-4}$\,J/mol\,K$^4$,   $\langle v_s\rangle \approx 3500$\,m/s is obtained.  Using $w=0.68$ mm and $t=0.20$ mm, we estimate the phonon thermal conductivity in the boundary limit $\kappa_{ph}^b/T\approx2.8\times 10^{-2}$\,W/K$^2$m at 0.1\,K, which is nearly four time larger than  the observed $\kappa/T$ at the lowest temperature.   Thus, $\kappa_{ph}$ is not  in the boundary limit even at the lowest temperatures. 

\newpage
\section{Analysis of Schottky anomaly}
\begin{figure}[h]
	\centering
	\includegraphics[width=\linewidth]{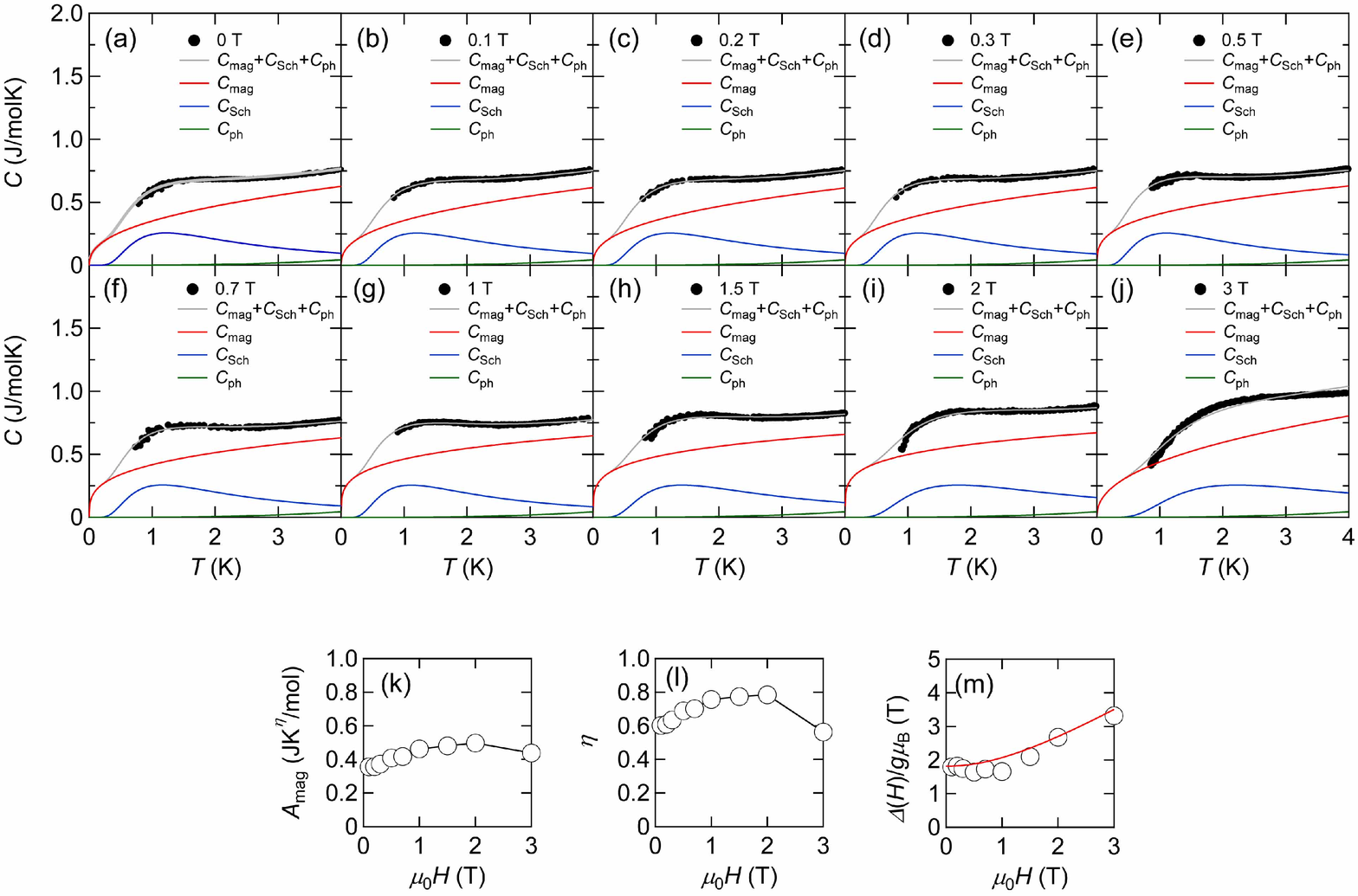}
	\caption{$C(T)$ vs. $T$  in (a) zero  and (b)-(j) in small magnetic fields  below 4\,K  where phonon contribution is negligibly small.    $C(T)$ is fitted by  the sum of $C_{Sch}$ (Eq.1) ,  $C_{mag}$ assuming power law temperature dependence of $C_{mag}=A_{mag}T^{1-\eta}$, where $A_{mag}$ and $\eta$ are constants, and $C_{ph}$.  Gray lines indicate the results of the fitting.  (k), (l) and (m) depict the field dependence of $A_{mag}$, $\eta$ and $\Delta(H)/g\mu_B$ obtained by the fitting.  The red line represents Eq.(2) with $H_0$=1.8\,T.}
\end{figure}

\newpage
\section{The entropy of $C_{mag}$}
\begin{figure}[h]
	\centering
	\includegraphics[width=0.6\linewidth]{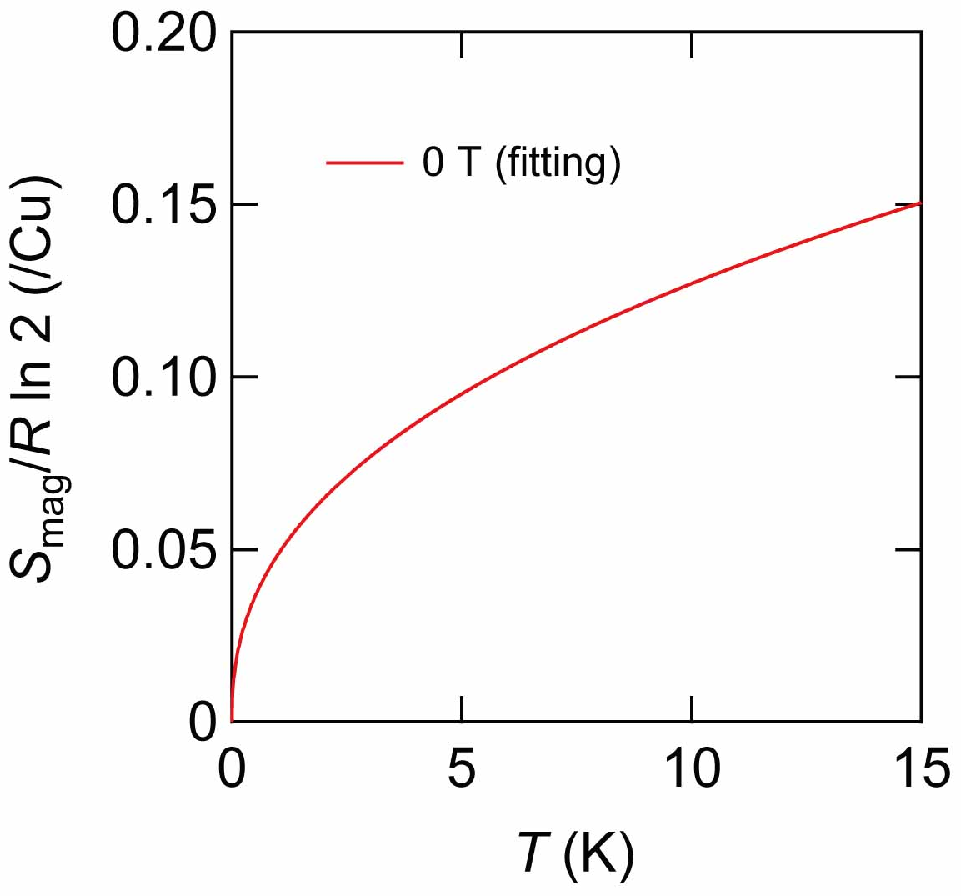}
	\caption{The entropy per Cu site normalized by $R\ln2$ is calculated by the formula $S_{mag}= (1/R \ln 2) \times \int^{T}_{0}\frac{C'_{mag}(T')}{T'}dT'$. $C'_{mag}=A_{mag}T^{1-\eta}$ $(A_{mag}=0.35, \eta=0.58)$ is the fitting function of $C_{mag}$ at 0 T.}
\end{figure}